# Time-resolved fast-neutron radiography of air-water two-phase flows in a rectangular channel by an improved detection system

Robert Zboray,[1] Volker Dangendorf,[2] Ilan Mor,[3] Benjamin Bromberger,[2] and Kai Tittelmeier,[2]

[1]*Paul Scherrer Institute, PSI Villigen, 5232, Switzerland*

[2]*Physikalisch-Technische Bundesanstalt (PTB), Braunschweig, 38116, Germany*

[3]*Soreq NRC, Yavne, 81800, Israel*

In a previous work we have demonstrated the feasibility of high-frame-rate, fast-neutron radiography of generic air-water two-phase flows in a 1.5 cm thick, rectangular flow channel. The experiments have been carried out at the high-intensity, white-beam facility of the Physikalisch-Technische Bundesanstalt, Germany, using an multi-frame, time-resolved detector developed for fast neutron resonance radiography. The results were however not fully optimal and therefore we have decided to modify the detector and optimize it for the given application, which is described in the present work. Furthermore, we managed to improve the image post-processing methodology and the noise suppression. Using the tailored detector and the improved post-processing significant increase in the image quality and an order of magnitude lower exposure times, down to 3.33 ms, have been achieved with minimized motion artifacts. Similar to the previous study, different two-phase flow regimes such as bubbly slug and churn flows have been examined. The enhanced imaging quality enables an improved prediction of two-phase flow parameters like the instantaneous volumetric gas fraction, bubble size and bubble velocities. Instantaneous velocity fields around the gas enclosures can also be more robustly predicted using optical flow methods as previously.

**I. INTRODUCTION**

Neutron imaging, in general, is a useful technique for visualizing low-Z materials (such as water or plastics) obscured by high-Z materials. For two-phase flows, thermal and cold neutron imaging have therefore found increasing application in the last three decades[1], mainly because they provide a better contrast for aqueous two-phase flows in a metallic piping in comparison to other techniques such as X- or gamma-rays. Especially interesting is the application for high-pressure, high-temperature two-phase flows in thick metal casings, e.g. in nuclear fuel bundle models. These studies aim to determine two-phase flow parameters with high spatial and/or temporal resolution in bundle geometries, which could then be used for fuel bundle optimization. A multitude of different flow regimes are encountered in a fuel bundle with convective boiling flows, ranging from low-gas fraction (bubbly flow) to very high gas-fractions (annular flow). For the latter flow regime in limited geometries (partial bundle), cold- and thermal-neutron imaging have been shown to be very useful[2]. However, when significant amounts of both low- and high-Z materials are present and full-bodied samples have to be examined, cold and thermal neutrons rapidly reach their limit and are not penetrating enough to be applicable. In such cases one can benefit from the high penetrating power of fast neutrons, which is a clear advantage for such cases, enabling higher-contrast imaging. This has been demonstrated comparing the performance of fast and thermal neutron imaging for the steady state measurement of gas distribution in a fuel bundle[3]. A fast-neutron imaging system is under development at the Paul Scherer Institute for fuel bundle studies and beyond[4].

Two-phase flow, in general, is a very rapidly changing process, requiring high-frame-rate imaging to capture its dynamics. In a previous study, we have already examined and demonstrated the feasibility of high-frame-rate, fast-neutron radiography for generic two-phase flows of relevance for fuel bundle studies[5]. However the detector setup used for the study turned out not to be optimal for this purpose. Therefore we have modified and optimized the detector and demonstrated its performance in another experiment, which is reported here. The experiments, just as for the previous study, have been carried out at the ion accelerator facility of the Physikalisch-Technische Bundesanstalt (PTB), in Braunschweig, Germany. For demonstration purposes, we have examined adiabatic, air-water two-phase flows at atmospheric pressures in a rectangular channel at different flow regimes, including bubbly, slug and churn flows. The instantaneous volumetric gas fraction, bubble sizes and bubble velocities and the velocity field around the bubble surfaces have been determined. We have managed to improve the image post-processing methodology to achieve a better noise suppression.

The next section presents a short description of the experimental setup highlighting the detector optimization and then the results obtained by the optimized detector and image post-processing are introduced.

## II. THE EXPERIMENTAL SETUP

The experiments have been carried out at the high intensity, white-spectrum, fast neutron beam line at PTB. The neutrons are produced by the d+Be reaction at an average energy of 5.5 MeV with a flux at the sample position of about $1.3e7$ cm$^{-2}$s$^{-1}$. The experimental setup including the beam line and the two-phase flow channel is described in detail in our previous paper[5]. We will only discuss here in detail the modifications in the detection system with respect to the previous work. The imaging detector used in the previous experiments was a third generation, multiple-frame Time-Resolved Integrative Optical Neutron (TRION) detector developed at PTB in the context of high resolution, energy-selective fast neutron resonance radiography[6,7]. It is based on a plastic-fiber scintillator screen and a two stage intensified CCD camera system. The components and layout of the detector are shown in Figure 1a. The fiber scintillator screen (BCF-12, produced by Crytur, fiber size 0.7 mm) has an active surface area of 200x200 mm$^2$ and is 50 mm thick, enabling a detection efficiency of 25.7 % at 6 MeV. Behind the bending mirror, a 120 mm lens focused the light on a position sensitive optical preamplifier (OPA), featuring a 40-mm Photek image intensifier (IMI), which intensifies the image from the scintillator screen and preserves the few-nanosecond fast timing property of the scintillator by using an E36 phosphor screen, which was important in the original neutron Time-of-Flight (TOF) applications. The intensified image from the OPA is split by a kaleidoscopic image splitter to a field of 3x3 sub-images. This image splitter is coupled through a lens to a 9-fold segmented IMI of which 8 segments can be gated independently. Each of the segments on the IMI photocathode views the scintillator screen and can acquire an image for an independently selectable time window with exposure times ranging down to 5 ns. The purpose of the 8-fold image splitting is to enable quasi-simultaneous recording of multiple images at different time slices, originally for TOF-sensitive imaging applications, in our case, for observation of a fast dynamic process. A CCD camera recorded all segments simultaneously on a large area CCD chip with 16 megapixels.

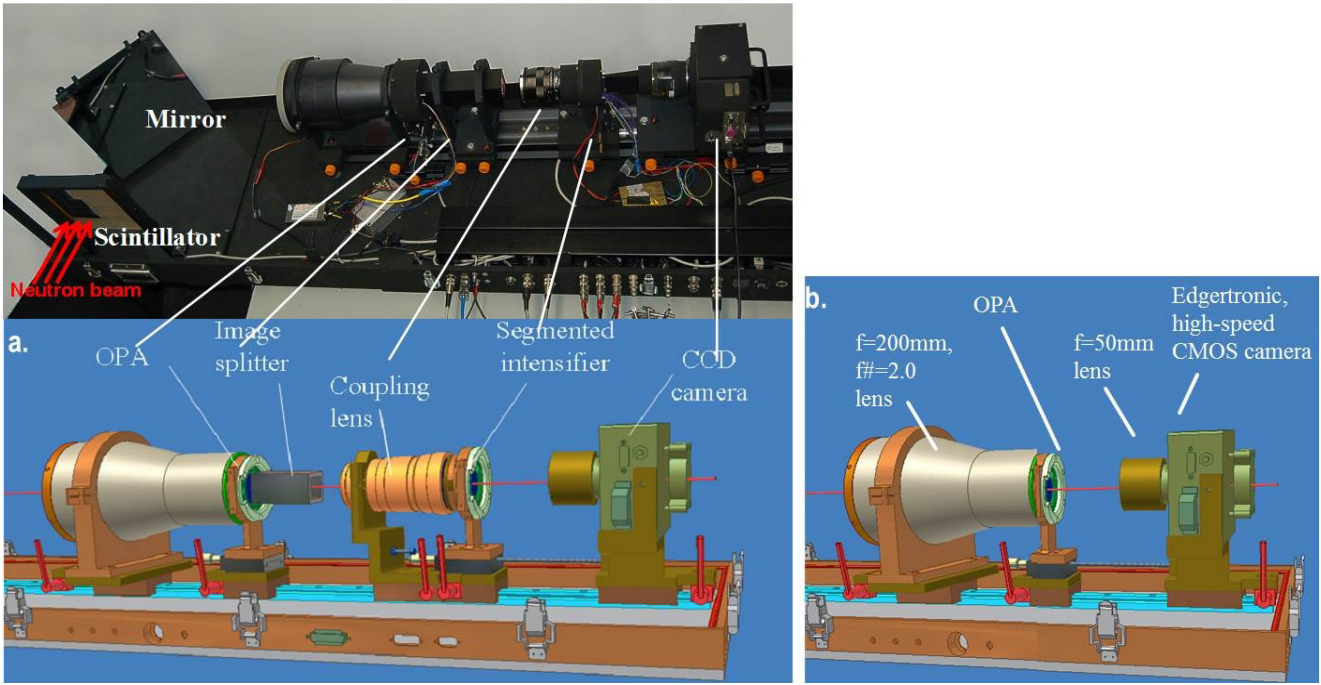

FIG. 1. (a) The generation III TRION detector developed at PTB and used for the present study. The flow channel (not shown) is placed in front of the scintillator screen in the neutron beam aligned with the vertical axis of the screen, perpendicular to the beam. (b) The optimized detector with a CMOS, high-speed camera (note that the scintillator screen and the mirror are not shown here).

Unfortunately, some adverse effects have been encountered using the TRION detector in the first experimental run. The two most severe are circular vignetting and intensity saturation in the images. Both are influenced by several factors, two important ones being the use of the fast but low light-yield E36 phosphor in the OPA and the presence of the image splitter, for details see our previous work[5]. These two light wasting stages required high gain in the OPA, which was realized by a triple multi-channel plate (MCP) intensifier stage. Despite using hot plates in the final amplification stage this setup is prone to saturation at high irradiation levels. These effects have been found only tolerable for the relatively low neutron fluxes in the pulsed beam experiments for neutron resonance imaging, for which TRION was originally developed. In our previous experiment and also here, the neutron flux is more than 20 times higher and therefore the saturation effects in the OPA cannot be any longer neglected. Therefore the detector has been modified and optimized for the experiments presented here. Compared to Figure 1a, in the modified detector the OPA is replaced by a standard 40 mm diameter IMI with a single MCP (multi-channel plate) amplification stage and a high light-yield P43 phosphor enabling to run the detector only with this single IMI. Consequently, the image splitter and the segmented IMI have been eliminated, solving the vignetting and the saturation problem. To enable a sufficient time resolution, the CCD camera has been replaced by a high-speed CMOS (made by Edgertronic, total 800x800 pixels) camera looking through a 50-mm lens directly on the OPA. The simplified and optimized detector is shown in Figure 1b except for the scintillator screen and the mirror which remained the same as in our previous work. Note that the use of the high-speed camera enables recording much longer image sequences with many thousands of frames and correspondingly longer observation times in contrast to the case with the 8-fold image intensifier in the original detector set-up. In addition to a more convenient and practical use of the detector, this enables a more detailed and better tracking of bubbles and their instantaneous parameters (size and velocity) as they move through the field of view.

The 15-mm thick, rectangular-shaped two-phase flow channel used to house the two-phase flow is described in our previous work, Zboray et al. (2014b). The positioning of the channel in front of the beam was very similar to that in our previous work, having a field of view (FOV) around 116x$116 mm$^2$. The bottom of the FOV starts around 1077 mm above the water inlet and 1027 mm above the air inlet. The channel was placed perpendicular to the neutron beam at a distance of L=617 mm from the source. The scintillator screen of the detector was at l=177 mm from the center of the channel.

**A. Spatial resolution and contrast transfer function**

As the channel positioning in the beam was very similar to our previous work, the spatial resolution is also very similar to that. We have used the same 3-steps tungsten mask, placing it directly on the front surface of the empty channel, to estimate the contrast transfer function (CTF) and the spatial resolution as in our previous work[5] and its geometry is described there. The image of the mask (normalized by the image of the empty channel) obtained for the optimized detector is shown in Figure 2a. The CTF, taken for the thickest step of the mask, is shown in Figure 2b implying a resolution using the usual limit value of 10 % CTF around 0.58 lp/mm. The spatial resolution is also estimated as the full width at half maximum (FWHM) of the derivative of the edge spread function (ESF). The latter is obtained by taking an average ESF over the thickest step of the 10-mm slit in the gray-scale image of the mask as indicated in Figure 2a and then it is fitted with a logistic function of the form:

$$ESF(x) = \beta_1 + \frac{\beta_2}{1+\exp[-\beta_3(\beta_4-x)]} \qquad (1)$$

where *x* is the space coordinate through the edge and a fit is illustrated in the inset of Figure 2b. FWHM values for different edges are given in the figure. They gradually increase for edges lying further from the image center due to blur effects we have analyzed in details in our previous paper[5], namely a parallax and a comet-tail-like" effect caused by the finite source size and by the thickness of the scintillator screen, respectively.

The pixel size of our images is 0.197 mm/pixel. Note that an internal mask on the camera has been applied making the raw image size to 592x592 pixels. Furthermore a 2x2-bining has been applied during image analysis doubling the aforementioned pixel size, which is still sufficient not to violate the sampling theorem given our actual spatial resolution.

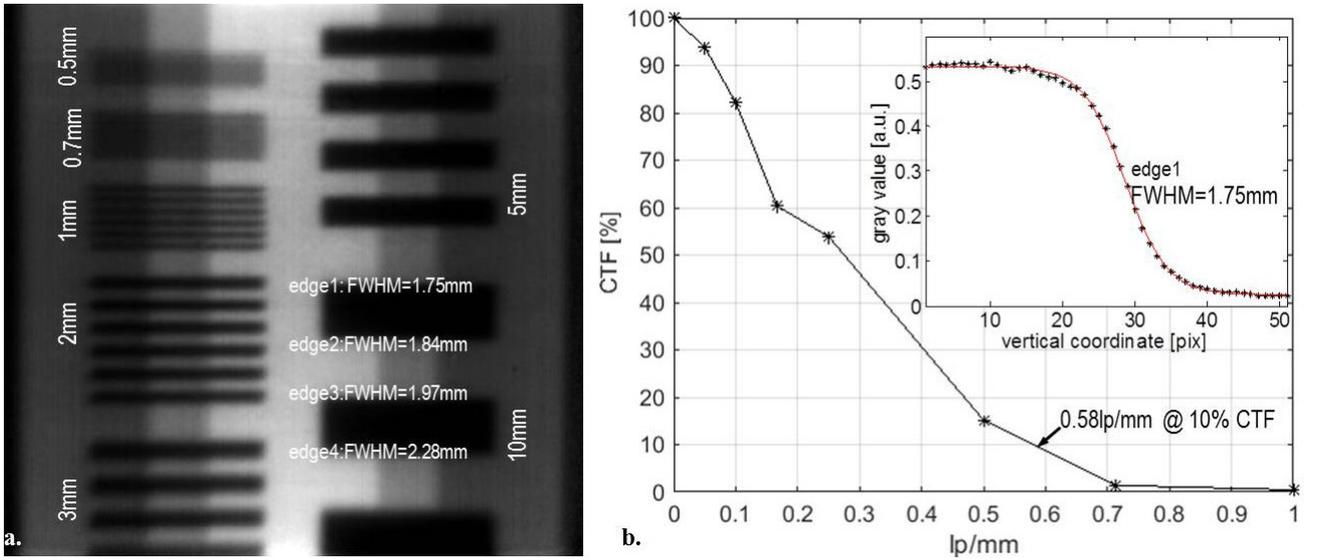

FIG. 2. (a) Flat-field corrected image of the attenuation of the tungsten CTF mask (the slit sizes are indicated). (b) The estimated CTF of the detector system taken at thickest step size of the mask (6, 12 and 20 mm). The inset in (b) shows a typical average edge response function (black asterisks) and the fitted logistic function (red). The edge response is taken over the edge1 indicated in (a).

## III. EXPERIMENTAL CONDITIONS

We have investigated different two-phase flow regimes from bubbly to churn flow by varying the water and the gas flow rates. All tests were carried out at an ambient temperature of approximately 20 °C and at atmospheric pressure (apart from some minimal pressure loss due to the flow through the channel). The two-phase flow can be characterized by the cross-sectional average volumetric gas and liquid fluxes (superficial velocities), *J*, defined as:
5

$$J_g = \frac{\dot{V}_g}{A}, J_l = \frac{\dot{V}_l}{A} \tag{2}$$

where $\dot{V}$ is the volumetric flow rate and $A$ is the area of the channel cross section. Furthermore, the gas volume flow fraction is defined as:

$$\varepsilon_{vf} = \frac{\dot{V}_g}{\dot{V}_l + \dot{V}_g} \tag{3}$$

Depending on the combination of the gas and liquid flow rates, the two-phase flow established in the channel shows different patterns ranging (in order of increasing gas volume flow fraction) from dispersed bubbly flow, through slug flow, to churn and annular flow. Cheng et al. give a comprehensive and recent overview of the different flow regimes in two-phase flows and their characteristics[8].

TABLE I. Matrix of experimental conditions.

| | Flow regime | Air flow [l/h] | Water flow [l/h] | $J_g$ [m/s] | $J_l$ [m/s] | $\varepsilon_{vf}$ [-] |
|---|---|---|---|---|---|---|
| Exp1 | Bubbly | 60 | 400 | 0.01 | 0.08 | 0.12 |
| Exp2 | Bubbly/Slug | 120 | 400 | 0.02 | 0.08 | 0.21 |
| Exp3 | Slug | 720 | 400 | 0.13 | 0.08 | 0.61 |
| Exp4 | Churn | 2280 | 400 | 0.41 | 0.08 | 0.83 |
| Exp5 | Bubbly | 60 | 600 | 0.01 | 0.12 | 0.08 |
| Exp6 | Bubbly | 60 | 800 | 0.01 | 0.17 | 0.06 |
| Exp7 | Bubbly | 60 | 1200 | 0.01 | 0.25 | 0.04 |
| Exp8 | Bubbly | 60 | 1400 | 0.01 | 0.29 | 0.04 |
| Exp9 | Bubbly | 120 | 1400 | 0.02 | 0.29 | 0.07 |
| Exp10 | Slug | 720 | 1400 | 0.13 | 0.29 | 0.31 |
| Exp11 | Slug/Churn | 2280 | 1400 | 0.41 | 0.29 | 0.59 |

Table I summarizes the four different experimental conditions and the parameters of the corresponding flow regimes. The latter were estimated based on phenomenological flow regime maps for narrow, rectangular channels with upward two-phase flow available in the literature[9]. Note that the transitions between different flow regimes are actually quite smooth and e.g. Exp2 lies in a transition region between bubbly and slug flow, whereas Exp11 between slug and churn flow.

## IV. EXPERIMENTAL RESULTS

To obtain the gas volume fraction in the channel, two reference (flat field) images are needed: one of the air-filled ($I_g$), and another of the water-filled channel ($I_w$). Based on that the instantaneous gas volume fraction, $\varepsilon$, at the image position *(x,y)*, averaged over the channel depth in beam *(z)* direction, can be given as:

$$\varepsilon(x,y) = -ln\left(\frac{I(x,y)}{I_w(x,y)}\right) / -ln\left(\frac{I_g(x,y)}{I_w(x,y)}\right) \qquad (4)$$

where *I* is the intensity of the image taken for the two-phase flow. Some typical instantaneous volumetric gas fraction ($\varepsilon$) distributions for different two-phase flow conditions are shown in Figures 3 to 5. Figure 3 shows a comparison for bubbly flow conditions with the results obtained in our previous experiments without optimizing the detector applying much longer exposure time. Note that the images from the previous experiment are processed using the same methodology as introduced in detail below to enable a meaningful qualitative comparison. The figures show the equivalent bubble diameter, $D_{eq}$, that is obtained from the estimated 3D instantaneous bubble volume $V_b$ as:

$$D_{eq} = \sqrt[3]{\frac{6V_b}{\pi}} \text{ with } V_b = d \sum_i \varepsilon(x_i, y_i)\, dx^2 \qquad (5)$$

where *d* is the channel thickness in beam direction, *dx* is the pixel size and the sum is taken over the pixels *i* contained in the bubble. Furthermore the positions of the center of mass (CM) of the bubbles and their instantaneous CM velocity based on the CM displacement between consecutive frames is also indicated on the figures.

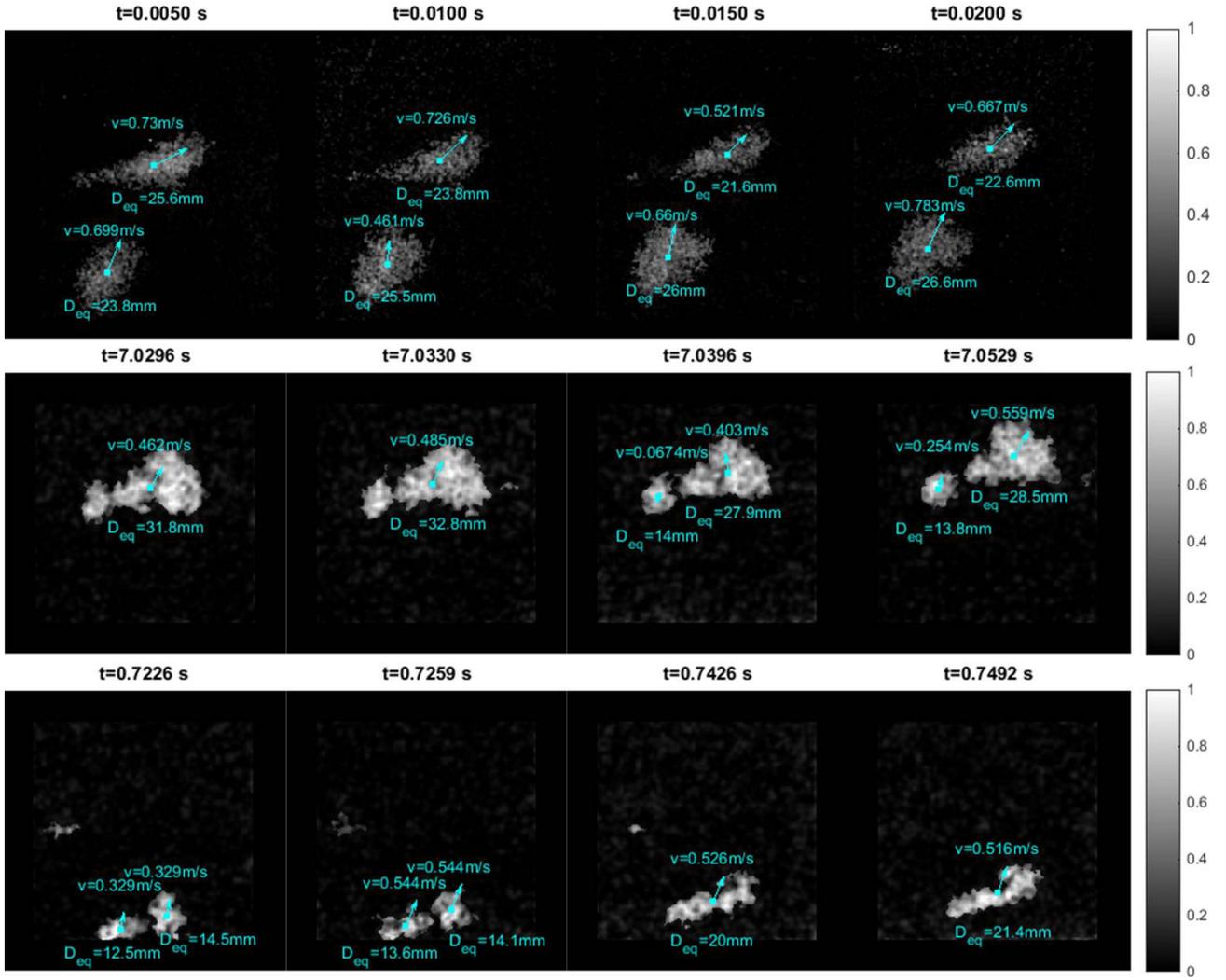

FIG. 3. Image sequences showing the instantaneous gas volume fraction for bubbly flows (Exp1). The upper row shows some results from the first experimental series with the unmodified detector using an exposure time of 30ms. Note that these images are processed using the same methodology as for the modified detector described in the present paper for a fair comparison. The middle and bottom rows depict results of the second experimental run using the optimized detector and an exposure time of 3.33ms. Note that the middle row shows the break-up of a large bubble, whereas the bottom row the coalescence of two smaller bubbles. The cyan dots indicate the center of mass of the bubbles and the arrows illustrate their instantaneous velocities obtained from the displacement of the centers of mass (arrow size is not to scale). The equivalent bubble diameter is also given in the figure. See Movie1 in the supplementary material[10] for a video of a section of Exp1.

Note that the Edgertronic high-speed, CMOS camera enables a range of different ISO sensitivity settings. After an initial tuning of the sensitivity setting, we have carried out all experiments once with ISO800 and also with ISO1600. In central image areas, no perceivable difference in image quality could be established, however at the image boundaries considerable higher noise is obtained using the ISO800 setting. Therefore all the images shown here are taken with ISO1600.

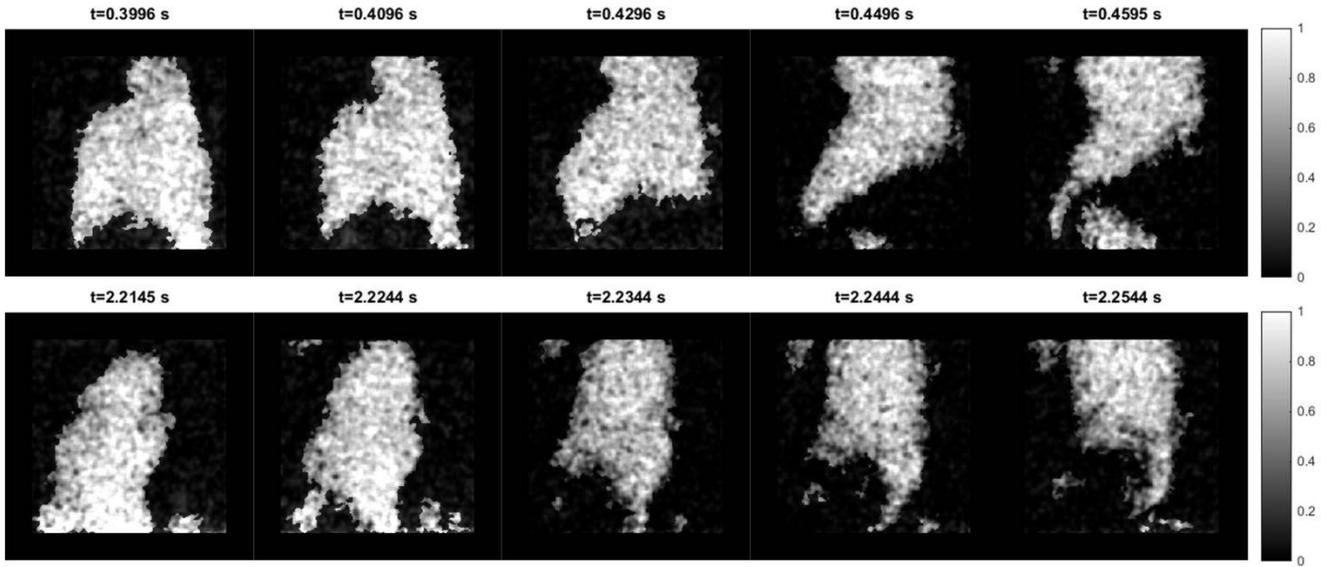

FIG. 4. Image sequences showing the instantaneous gas volume fraction for slug flow (Exp3) captured with an exposure time of 3.33ms. See Movie2 in the supplementary material[10] for a video of a section of Exp3.

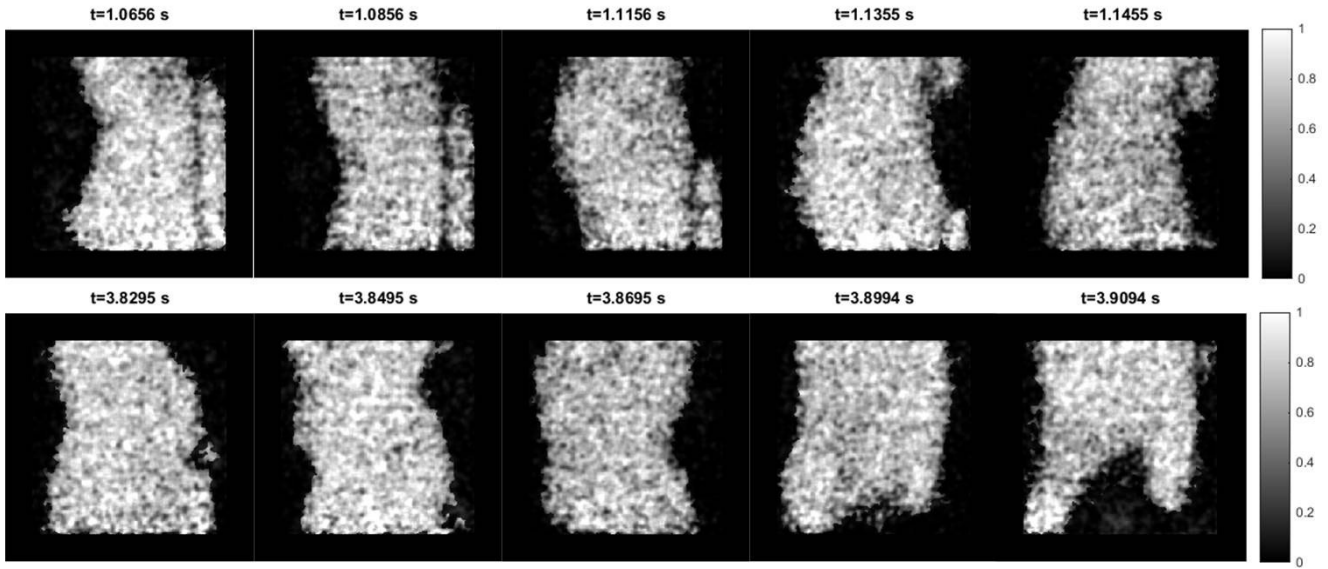

FIG. 5. Image sequences showing the instantaneous gas volume fraction for churn flow (Exp4) captured with an exposure time of 3.33ms. See Movie3 in the supplementary material[10] for a video of a section of Exp4.

To reduce shot noise and, in general, to filter the images of unwanted features, several post processing steps are taken, which we describe here in detail. First a 3x3 pixel spatial median filter is applied and secondly an anisotropic diffusion based filter as proposed originally by Perona and Malik[11] that we already used and explained in our previous work[5]. Then separately, the true bubble contours (areas) are identified based on motion continuity. For this purpose moving averaging over three consecutive frames in time is done centered on the actual instantaneous frame, on the previous and the next frames. For all three moving averaged images a threshold filter is applied based on their pixel intensity distribution. It is found that a very significant part of the noise is robustly removed while bubbles are preserved if, for bubbly flow all pixels below the 0.9 quantile value, for slug flow below the 0.7 quantile value and for churn flow below the 0.55 quantile value are set to zero. The filtering is needed to facilitate the next processing step of bubble identification based on a connected component labelling algorithm. For this the *bwlabel* function of MATLAB[12] is used. The five largest connected areas in all the three moving-averaged images are taken as potential bubble contours (*bwlabel* delivers a black and white, BW, image) and in the next step are tested for being a true bubble. We confined the bubble identification algorithm to the five largest bubbles as experience shows that it is extremely rare to find more than five bubbles at the same time on our images. Each of the five potential bubbles on each three moving-averaged images are taken into a separate BW image of the same size as the original image. Then the image of each bubble candidate in the first frame, $I_1$ is paired with the images of all the bubble candidates in the two consecutive frames, e.g. $I_2$, and the following measure is taken as a base of bubble identification:

$$\sum_{m,n=0}^{M-1,N-1}(cor(I_1,I_2)-cor(I_1,I_1))^2 / \sum_{m,n=0}^{M-1,N-1}cor(I_1,I_2) / \sum_{m,n=0}^{M-1,N-1}cor(I_1,I_1) \qquad (6)$$

where

$$cor(I_1,I_2)(k,l) = \sum_{m,n=0}^{M-1,N-1} I_1(m,n)I_2(m-k,n-l) \qquad (7)$$

is the (cross)correlation matrix between the image matrices $I_1$ and $I_2$. In an ideal case, when the bubble preserves exactly its shape between two consecutive images the measure in Eq. (6) is zero. For slight deformation, as is the case in practice, it still assumes a very small value close to zero. In our images, we found that this value is always clearly below 0.03\% for matching bubbles in consecutive images, therefore, it was used as threshold for bubble identification. For non-matching bubbles or noise this value is always at least an order of magnitude higher. Note, that we have also evaluated other test conditions for bubble identification, than the above correlation-based measure, like similarity of bubble shape (eccentricity) and size, etc. However their performance was inferior compared to Eq. (6). The inverse BW image of the contours of the identified bubbles is then used to suppress every pixel outside of the bubbles to one quarter of their values in the actual image frame, while the pixels inside the contours are left at their original brightness values. Figure 6 illustrates by an example the efficiency of the image filtering and post-processing method comparing the raw camera image, the non-filtered but flat-field compensated image (cf. Eq. 4) and the filtered, post-processed image.

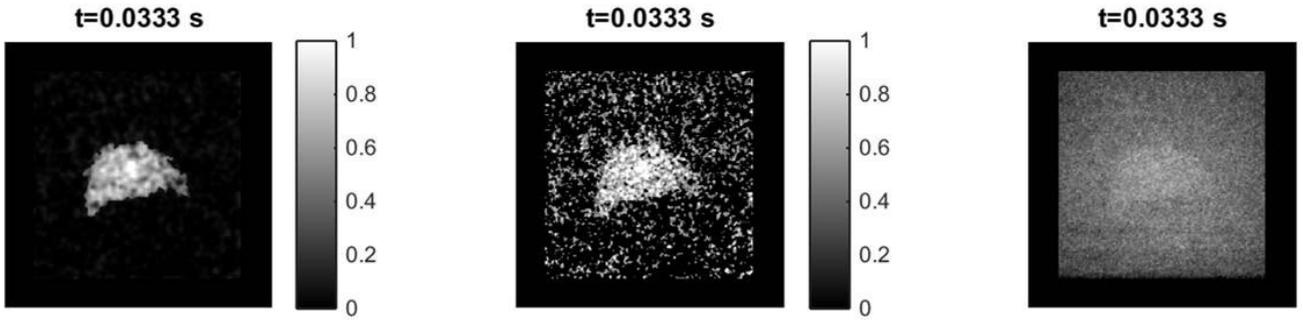

FIG. 6. Illustration of the image post processing and filtering method: (a) filtered and flat-field compensated (volume fraction scaled), (b) flat-field compensated image without filtering and (c) raw camera image (grey scale in arbitrary units). Exp1, exposure time: 3.33ms.

Note that it is necessary to use at least three consecutive image frames for the bubble identification to be able to handle cases of bubble coalescence and/or break up between consecutive frames. Using more than three consecutive frames would, however, complicate the processing and increase computational effort. Note also, that for churn flow (Exp4 and Figure 5), the above post processing methodology seems to work as well, although no dispersed bubbles are present, rather a large-scale, interconnected gas enclosure.

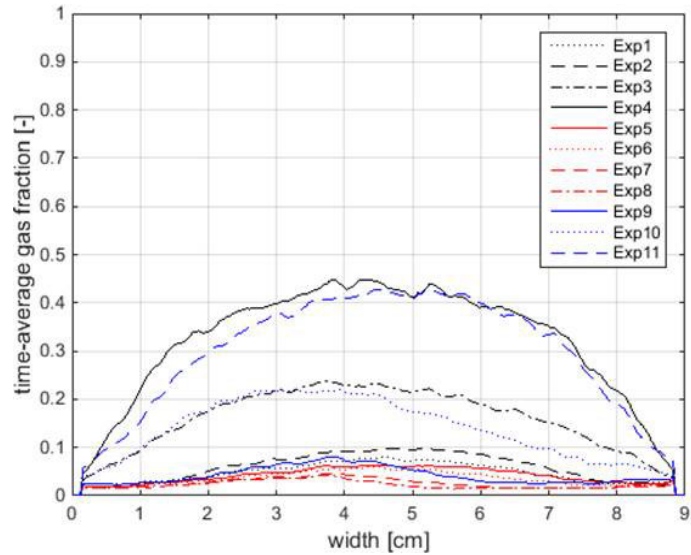

FIG. 7. The time-averaged lateral gas volume fraction profile over the channel for all experimental conditions obtained for 3.33ms exposure time. The curves are obtained by averaging the volume fraction vertically over the FOV.

Figure 7 shows time-averaged, lateral gas-volume-fraction profile over the channel for all experimental conditions obtained for 3.33ms exposure time. The volume fraction images are averaged vertically over the FOV to obtain the lateral profiles. A slight lateral asymmetry of some profiles is likely due to the relatively short total measurement time (10 s) and the tendency of the large-scale flow structures to fluctuate laterally in the broad (9 cm) channel evidenced by post-test visual observations.

Figure 8 shows the instantaneous, equivalent bubble size distribution and the instantaneous CM bubble velocity distributions. The distributions are somewhat noisy due again to the relatively short measurement times, but some trends are clearly observable. Moving from dispersed bubbly flows (Exp1,5,6,7) to bubbly/slug (Exp2), slug (Exp3,10) and slug/churn (Exp11) flows, there is clearly a higher contribution at large equivalent bubble sizes (above 25-30 mm) as expected based on the flow patterns in Figures 3 to 5. For bubbly flows there is an extremely low contribution above about 28 mm. A typical bimodal-type size distribution can be observed in case of bubbly flows, a smaller second hump is present in the distribution around 20 mm beside the peak at small sizes (around 5 mm). This is most pronounced for Exp1 and is well known in the literature[13] and is due to bubble coalescence occurring over the large distance from the gas injection to the FOV (more than 40x$D_h$, where $D_h=4A/P$ is the hydraulic diameter of the channel $P$ being its wetted perimeter). Furthermore, for bubbly flows with constant air but with increasing water flow (Exp1, Exp5 to 8) Figure 8 shows a gradually decreasing maximal bubble size and also the peak at small sizes shifts to slightly lower values. This is likely due to lower initial bubble sizes caused by the higher shear forces by the water acting on the gas at the air injection orifices at larger water flow rates. This causes correspondingly also smaller coalesced bubble sizes.

Regarding the instantaneous CM bubble velocity distributions, the major contribution is up to around 0.5-0.6 m/s, beyond that, the curves start declining, although in a different manner, for all the experiments having a very little contribution above 0.9 m/s. This range of bubble velocities corresponds roughly to the sum of the liquid superficial velocities (see Table I) and the typical bubble terminal velocities. The latter increase proportional to the square-root of the equivalent diameter for large distorted bubbles[14], and should vary between 0.2-0.4 m/s for $D_e$=10-35 mm.

Note that the bubble size and the CM velocity is only taken for bubbles appearing fully inside the FOV. Each experiment has also been repeated for two different exposure times namely 5 ms and 3.33 ms. Note, though it is not shown in the paper for brevity, that no major differences have been found in the results for the two exposure times regarding e.g. bubble sizes and velocity distribution. Nevertheless, we show results obtained with the shorter exposure time as any eventual motion artefact must be smaller for those. Regarding the effect of motion blur, we can make some coarse estimate as follows: taking 3.33 ms exposure time and 0.4-0.5 m/s bubble velocity results in a motion blur of about 1.5 mm. This together with our motionless spatial resolution from Figure 2 results in roughly $\sqrt{1.75^2 + 1.5^2}$ =2.3 mm effective spatial resolution in the image center and $\sqrt{1.75^2 + 2.3^2}$ =2.9 mm towards the image periphery. Therefore the smallest resolvable bubble size is somewhere between 3 and 4 mm.

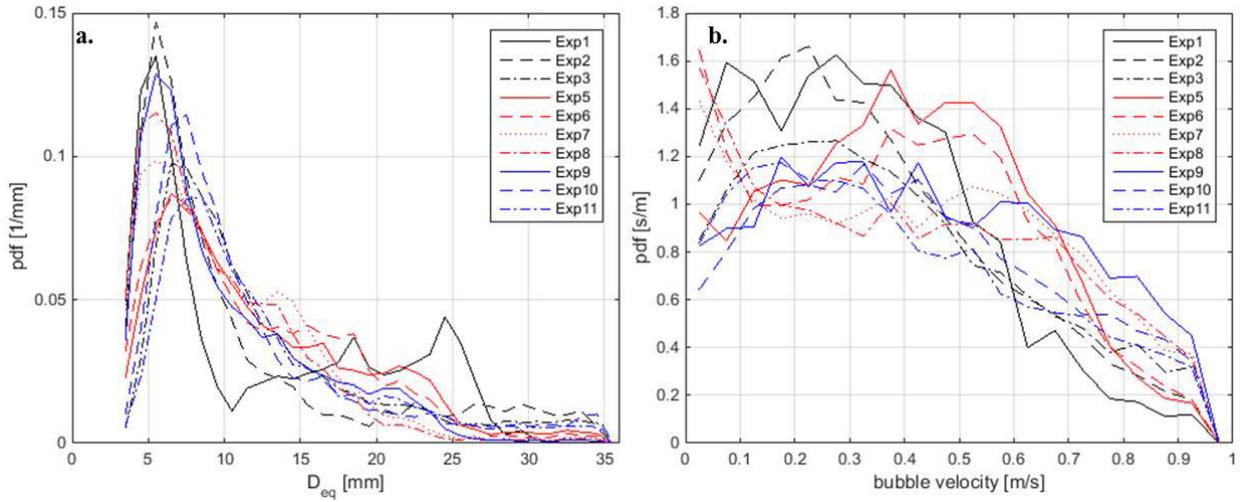

FIG. 8. (a) The instantaneous equivalent bubble size distribution and (b) the instantaneous CM bubble velocity distribution based on the instantaneous images for 3.33ms exposure time. Results are shown for all experimental conditions except for Exp4, being a churn flow, where the interpretation of the gas enclosures as dispersed, individual bubbles is not anymore possible.

Besides the instantaneous CM bubble velocities, similar to our previous work, we also attempt here the use the sequence of instantaneous gas-fraction images to resolve the velocity fields around and in the moving bubbles and gas enclosure. This, while complementing the CM velocity estimate for very dispersed (bubbly) flows, is certainly needed for e.g. churn flows where no well-separated gas enclosures occur but a very complex, fragmented interface between the two phases occur (see Figure 10). We try to apply for this the optical flow (OF) method just as in our previous study, however instead of using the in-house implementation of the algorithm, we tried two robust implementations of the OF method available in the latest version of MATLAB 2014b[12]. One of them is using the original Horn-Schunk (HS) formulation[15], we also used previously[5]. The second one is based on the Lucas-Kanade (LK) method[16].

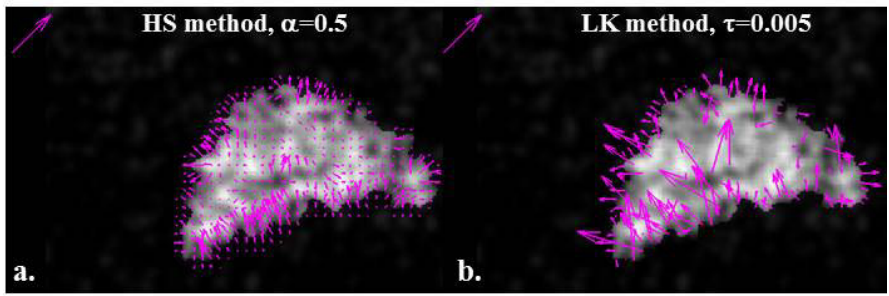

FIG. 9. Instantaneous, two-dimensional gas-velocity field estimation in and around bubbles (Exp1) using optical flow methods. Two different formulations are tested: the Horn-Schunk (a) and the Lucas-Kanade (b) method. The magenta arrows in the left-upper corners correspond to a velocity magnitude of 0.5 m/s. Note that only every fourth vector in the velocity fields is depicted for better visibility.

Both algorithms work on the premise that the pixel-wise image intensity, $I$, at coordinates $x,y$, is conserved from one image to the next in the sequence, at least up to $1^{st}$-order making it ideal for small displacements:

$$\frac{DI(x,y,t)}{Dt} = 0 \tag{8}$$

resulting in:

$$I_x u + I_y v + I_t = 0 \tag{9}$$

where $I_x$, $I_y$, $I_t$ are the spatio-temporal image brightness derivatives and $u,v$ are the velocity components in the $x,y$ directions, respectively. As the problem is ill-posed in this form having one equation for the two velocity components, a smoothness constraint is added in the HS formulation making it an optimization problem to minimize the goal function:

$$L = \int \left[ \left(I_x u + I_y v + I_t\right)^2 + \alpha^2 (|\nabla u|^2 + |\nabla v|^2) dx dy \right] \tag{10}$$

where the parameter $\alpha$ weights the global smoothness term and its value influences the velocity estimate. The LK method divides the original image into smaller (nxn pixels) sections and assumes a constant velocity within each section, thereby making the problem over determined and thus eliminating the need for the smoothness criterion. Then it performs a weighted least-square inversion of Eq. 9 for $u$ and $v$. The division of the image can be done by one-pixel shifted, overlapping sections thereby retaining the original pixel resolution for the velocity field or iterative pyramidal refining of the velocity field is applied in practice as well. For the LK method, there exist a $\tau$ parameter[12], which is compared to the eigenvalues of the least-square solution matrix for $u$ and $v$ and the solution is only taken if both eigenvalues are larger than $\tau$. If both are smaller than $\tau$ the result is 0. If one is larger other is smaller than the gradients Eq. 9 are normalized to calculate $u$ and $v$. This is done to reduce noise and false vectors and the LK method indeed produces no less vectors in regions with low texture (inside the bubbles) than the HS method. This is illustrated in Figure 9. In general the lower $\alpha$ or $\tau$ are, the noisier are the vector fields. Examining the change in the mean and RMS values of the two velocity components by varying $\alpha$ and $\tau$ for our typical bubble images, we have concluded that $\alpha=0.5$ and $\tau=0.005$ are relatively robust choices (see Figure 9). Furthermore, one can observe that the HS method tends to fill out the velocity field with vectors further beyond the edge of the bubble than the LK does. This is obviously due to the global smoothness criterion and this tendency we have already observed in our previous work[5]. In general, again due to the smoothness criterion, the HS tends to produce smoother and fuller velocity fields. It is also producing more accurate results compared to the LK method according to the Middlebury OF benchmark[17]. In regions of high texture, large image gradients (bubble edges) both methods produce similar results, the relative difference being around 10-15 %.

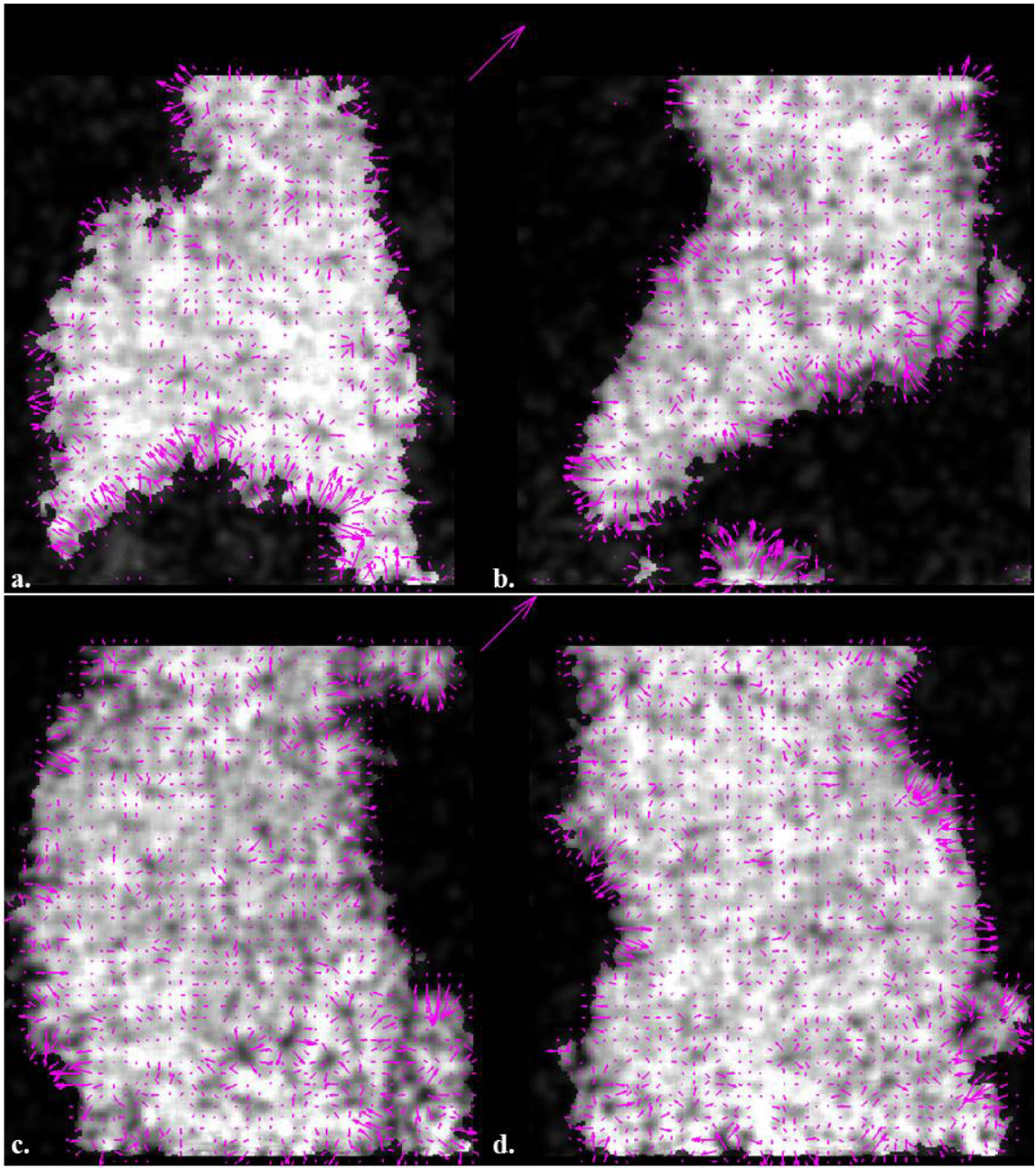

FIG. 10. Illustrations of instantaneous, gas-velocity field estimations in and around gas enclosures for (a), (b) slug flow (Exp3); and for (c), (d) churn flow (Exp4) using optical flow methods using the HS formulation. The magenta arrows at the top correspond to a velocity magnitude of 0.5 m/s. Note that only every fifth vector in the velocity fields is depicted for better visibility.

The performance of the HS method with the optimized value of $\alpha$ for slug and churn flows is illustrated in Figure 10. Note that inside bubbles and gas enclosures the frame-to-frame image brightness variations are due also to some extent to noise. The OF algorithm is not able distinguish that contribution from actual bubble deformation and motion, however these contributions are usually quite small, due partly to the smoothness criterion, as illustrated on the figures and if needed could easily be removed by threshold filtering.

## V. CONCLUSIONS

We have shown the feasibility of time-resolved, fast neutron radiography of generic two-phase flows in a 15-mm thick channel. A white-spectrum, intense fast-neutron beam in conjunction with optical, fiber scintillator-screen-based fast neutron imaging detectors, developed originally for fast neutron resonance radiography, have been employed. The detector has been simplified and optimized for the present application featuring a single image intensifier and a high-speed CMOS camera also enabling much longer observation times with many thousands of frames in a sequence and thus a more convenient and practical use of the detector and the available neutron beam. We managed to improve the image post-processing methodology, which in turn substantially improved the image quality and the noise suppression. It was shown that a multitude of two-phase flow parameters of practical interest can be derived from these measurements such as: the instantaneous gas volume-fraction distribution, bubble size distributions, distribution of the center-of-mass bubble velocity. All of these have been achieved at exposure times down to 3.33 ms per frame, an order of magnitude lower than in our previous study[5], thereby minimizing motion artifacts. Bubbles with diameters down to 3-4 mm can be resolved. We have applied the optical flow method on the time-resolved images to obtain instantaneous, two-dimensional velocity fields in and around bubbles and gas enclosures. Two robust implementations of the method have been tested and we managed to achieve an improved performance for the velocity field prediction compared to our previous work as well.